\documentclass[aapm,mph,superscriptaddress,reprint,longbibliography]{revtex4-2}

\usepackage{xcolor,soul}
\usepackage{acronym}
\usepackage{amsmath,amssymb}
\usepackage{isotope}
\usepackage{graphicx}
\usepackage{siunitx}
\usepackage{stackrel}

\newcommand*{\pd}[1]{\partial_{#1}}
\newcommand*{\dd}[1][{}]{\mathrm{d}#1}
\renewcommand{\vec}[1]{\mathbf{#1}}
\newcommand{\e}{\ensuremath{\mathrm{e}}}
\let\div\relax
\DeclareMathOperator{\div}{div}
\DeclareMathOperator{\grad}{grad}

\begin{document}

\newacro{DICOM}{Digital Imaging and Communications in Medicine}
\newacro{CT}{computed tomography}
\newacro{MRI}{magnetic resonance imaging}
\newacro{CA}{contrast agent}
\newacro{VOI}{volume of interest}
\newacro{VIF}{vascular input function}
\newacro{EES}{extracellular extravascular space}
\newacro{DCE}{dynamic contrast enhanced}
\newacro{mDIXON}{modified Dixon}
\newacro{eTHRIVE}{enhanced T1-weighted high-resolution isotropic volume examination}
\newacro{TM}{Tofts model}
\newacro{3CM}{three-compartmental model}
\newacro{Eovist}[Gd-EOB-DTPA]{gadoxetate disodium}

% \preprint{}

\title{Multi-compartmental modeling for Gd-EOB-DTPA using sparse human DCE-MRI data}
% short title: Multi-compartmental DCE-MRI modeling of sparse data

% \author[1]{Christian Velten}{\orcid{0000-0002-0081-873X}}
% \author[1,2]{Megi Gjini}
% \author[1,2]{Patrik Brodin}
% \author[1,2]{Wolfgang A. Tome}
\author{Christian Velten}
\author{Megi Gjini}
\author{N. Patrik Brodin}
\author{Wolfgang A. Tom\'{e}}
\affiliation{Department of Radiation Oncology, Montefiore Medical Center, Bronx, NY, USA}
\affiliation{Institute for Onco-Physics, Albert Einstein College of Medicine, Bronx, NY, USA}

% \authormark{VELTEN \textsc{et al.}}
% \address[1]{\orgdiv{Institute for Onco-Physics}, \orgname{Albert Einstein College of Medicine}, \orgaddress{\state{NY}, \country{USA}}}
% \address[2]{\orgdiv{Department of Radiation Oncology}, \orgname{Montefiore Medical Center}}
% \corres{Christian Velten, This is sample corresponding address. \email{cvelten@montefiore.org}}
% \presentaddress{This is sample for present address text this is sample for present address text}
% \finfo{This work was partially supported by \fundingAgency{National Science Foundation} grant \fundingNumber{DMS-2014626}}

\begin{abstract}
	\noindent\textbf{Purpose:} To derive kinetic equations for multi-compartmental contrast agent distribution from first principles and apply it to two and three compartments for Gd-EOB-DTPA using low time resolution human liver DCE-MRI data.

	\noindent\textbf{Methods:} The continuity and diffusion equation were combined and used to derive a general form for differential equations governing multi-compartmental particle exchange. They were applied to two (equivalent to the Tofts model) and three compartments. Both models were fit to human DCE-MRI data with low temporal resolution and three compartment model's parameters' implications are discussed.

	\noindent\textbf{Results:} The model derived for two compartments is shown to be equivalent with the Tofts model. Using reasonable biological and physical assumptions an analytical solution for the three compartment model is obtained. The three compartment model was able to fit all Gd-EOB-DTPA DCE-MRI data whereas the Tofts model did not. Differences were increased for cases with large as well as rapid uptake of contrast.

	\noindent\textbf{Conclusion:} We demonstrated the ability to fit sparse Gd-EOB-DTPA DCE-MRI data using a three compartment model derived from first principles whose parameters can be used to help quantify overall and regional liver function.
\end{abstract}


\maketitle

% ======================
% BODY-BEGIN
% ======================

\section{Introduction}\label{sec:Introduction}

Volumetric imaging can be used to quantify the spatial and temporal contrast agent voxel density, i.e. the amount of signal producing material within the volume of one voxel. \acs{MRI} is extensively used for \acs{DCE} imaging due to its non-ionizing nature and superior soft tissue contrast compared to \acs{CT}. One of the most prominent and most widely used kinetic modeling approaches is that described by \citeauthor{Tofts1999a} that describes the exchange of contrast material between two compartments, blood plasma and \ac{EES} based on an \ac{VIF}.\cite{Tofts1999a,Sourbron2011} \citeauthor{Sourbron2013} have described generalizations of these classical models and alternative approaches to kinetic modeling.\cite{Sourbron2013} Extensions of the \ac{TM} described in the literature include the inclusion of spatial diffusion from adjacent tissues into the \ac{VOI}.\cite{Cantrell2017}

\acs{MRI} \acp*{CA} can be classified into non-specific agents which distribute only in vascular and extracellular spaces and into tissue-specific agents which show cellular uptake. For example, \ac{Eovist}, also known under its trade name {Eovist\textsuperscript\textregistered}, is a liver-specific \ac{CA} designed to be taken up by hepatocytes. Its characteristics have been described elsewhere~\cite{VanBeers2012,Choi2016}. Briefly, the \ac{CA} is transported into hepatocytes through organic anion transporting polypeptides OATP1B1 and OATP1B3 and subsequently excreted into bile via the multidrug resistance protein 2 (MRP2). A differential between the transport rates of OATP1B1/B3 and MRP2 leads to net \ac{CA} retention in healthy hepatocytes. In contrast, hepatocellular carcinomas tend to have decreased OATP1B1/B3 and increased MRP2 expression, leading to decreased uptake and quicker release of \ac{CA}. These features require the use of models taking into account retention of contrast agent over the course of minutes or hours, leading to different imaging behavior compared to simple flow- or permeability-limited contrast agents and models, like the \ac{TM}. 

The use of multi-compartmental models has been described previously~\cite{Forsgren2014,Georgiou2017}, where \citeauthor{Forsgren2014} proposed multiple models with up to six free fit parameters and \citeauthor{Georgiou2017} had access to high-resolution temporal data. Routine \acs{DCE}-\acs{MRI} exams, however, may only contain up to three \textsl{dynamic} (acquired within the first five minutes following \ac{CA}) and one or two \textsl{delayed} images (acqcuired twenty to thirty minutes after \ac{CA} injection).
Similar to previous previously published work, we derived the differential equations describing the kinetics of multiple compartments using the continuity and diffusion equations, which when applied to two compartments are shown to be equivalent to the \ac{TM}. Finally, we applied it to three compartments, compared the performance of both models in a sample of \acs{DCE}-\acs{MRI} scans with sparse data, and gave some interpretation of the fit parameters as they relate to OATPB1/B3 and MRP2 expression.

The approach presented in this work can be used to model multi-compartmental kinetics where most cases with more than two to three compartments may need to be solved numerically. Finally, while the kinetics of \acs{MRI} \acp*{CA} are used as an example this approach can be applied to both, other agents with specific or unspecific behavior and to model kinetics of inter- or intracellular distribution.

% \cite{Tofts1999a}

\section{Methods}\label{sec:Methods}

A \ac{VOI}, may be composed of different compartments that exchange particles with one another and with an external reservoir. Measurements of signal intensity are not able to distinguish between signal emanating from one compartment over another but relate to the overall concentration of particles in the \ac{VOI} (app.\;\ref{sec:derivation}, eqn.\;\ref{eqn:app_signalpropto}),
\begin{align}
	\text{Signal} \propto \phi_{V} = \dfrac{\Phi_V}{V} = \dfrac{1}{V}\,\sum\limits_{i}\phi_i\,V_i = \sum\limits_{i}\nu_i\,\phi_i
\end{align}
The differential equation describing the exchange of particles between compartment $i$ with a number of compartments $\{j\}$ has been derived in appendix~\ref{sec:derivation} and, assuming particle number conservation and appropriateness of diffusion processes, is given by (eqn.\;\ref{eqn:app_final_ode})
\begin{align}\label{eqn:compartmentkinetics}
	\nu_i\,\pd{t}\phi_i(t) = \sum_{j}\left[ k_{i\leftarrow j}\phi_j(t) - k_{i\rightarrow j}\phi_i(t) \right]
\end{align}
Assuming that one compartment $i$ acts as initial supplier of particles, thus \hbox{$\phi_{j\neq i}(t_0 \leq 0)=0$}, the overall concentration  $\phi_V$ may also be assumed to be a convolution of said input compartment and response functions $R_{j}$,\cite{Georgiou2017}
\begin{align}\label{eqn:responsefcn}
	\phi_V = \phi_{i} * R = \sum\limits_{j\neq i} \phi_{i} * R_j
\end{align}

In the following sections we apply these equations to two compartments to show its equivalence with the \ac{TM} and three compartments to model \ac{CA} uptake and retention. For both cases, response functions are also calculated by applying Laplace transforms to both equations \ref{eqn:compartmentkinetics} and \ref{eqn:responsefcn} under the initial conditions that all but the initial supply compartment $i$ are empty, i.e. \hbox{$\phi_{j\neq i}(t_0\leq 0)=0$}.

\subsection{Two Compartment Model}\label{sec:TwoCompartmentModel}
Consider two points, or compartments, $A$ and $B$, where only a part of the latter, $B^\prime$, with volume $V_{B^\prime}=\nu\,V_B$, can exchange particles between them. We will assume that we know the time dependent particle density in compartment $A$ and will thus write for compartment $B^\prime$:
\begin{align*}
	\nu\,\pd{t}\phi_{B^\prime}(t) &= k_{B^\prime A}\left[\phi_A(t)-\phi_{B^\prime(t)}\right] \\
	\pd{t}\phi_{B}(t) &= k_{BA}\left[\phi_A(t)-\dfrac{\phi_{B}}{\nu}\right]
\end{align*}
Where we used $\phi_{B^\prime}=\nu^{-1}\phi_{B}$, $\nu:=\nu_{B^\prime}$, $\nu_{B}=1$, $k_{B^{\prime}A}=k_{BA}$, and that the absolute change in particles in compartments $B$ and $B^\prime$ is the same, namely that $\pd{t}\Phi_{B}(t) \equiv \pd{t}\Phi_{B^\prime}(t)$. This is because the latter is fully contained in the former and is the only volume able to exchange particles; there are no particles contained in $B\setminus B^\prime$. Solving for $R(t)$ and $\phi_B(t)$ yields
\begin{align*}
	R(t) &= k_{BA}\,\exp\left\{-\dfrac{k_{BA}}{\nu}\,t\right\} \\
	\phi_B(t) &= k_{BA}\int\limits_{0}^{t}\dd{\tau}\,\exp\left\{-\dfrac{k_{BA}}{\nu}(t-\tau)\right\}\,\phi_A(\tau)
\end{align*}
The differential equation and its solution above can easily be identified to be equivalent to the generalized kinetic model described by \citeauthor{Tofts1999a},
\begin{align*}
\pd{t}C_{t} &= \ensuremath{K^{\mathrm{trans}}}\left(C_p - \dfrac{C_t}{\nu_{e}}\right) = \ensuremath{K^{\mathrm{trans}}}\,C_p - k_{\mathrm{ep}}\,C_t \\
\ensuremath{K^{\mathrm{trans}}} &:= k_{BA} = \dfrac{1}{V}\left(P\,S\right)_{B\leftrightarrow A} \\
k_{\mathrm{ep}}&:=\nu^{-1}\,\ensuremath{K^{\mathrm{trans}}} = \dfrac{1}{\nu\,V}\left(P\,S\right)_{B\leftrightarrow A}
\end{align*}
where $P$ is the permeability of surface $S$ across which diffusion takes place (see app.\;\ref{sec:derivation} for details).

This is not surprising as the \ac{TM} describes the exchange of contrast agent between plasma and \acf{EES} in tissue with no retained uptake in the tissue itself. The dynamics of the contrast distribution depend only on supply in the vasculature (blood pool, $\phi_A$) and the particle transfer rates between vasculature and \ac{EES} (\ensuremath{K^{\mathrm{trans}}}~\& $k_{\mathrm{ep}}$).
Since there is always exchange between both compartments, $\nu\in\left(0,1\right]$ and thus, $k_{\mathrm{ep}}\geq\ensuremath{K^{\mathrm{trans}}}$, this model cannot be used to describe long-term contrast retention or delayed release of contrast.
%However, by adding a multiplication factor $\lambda$ with $\lambda\in\left[0,1\right]$ to $\phi_B/\nu$ in the differential equation above, we can adjust the fraction of the density in $B$ that affects the diffusion balance through the gradient between $A$ and $B$. This leads to the same equations with an additional factor $\lambda$ in the exponential, such that $k_{\mathrm{ep}}=\frac{\lambda}{\nu}\ensuremath{K^{\mathrm{trans}}}$. For data modeling one may set $\tilde{\nu}:=\nu/\lambda$ with $\tilde{\nu}\in\left(0,\infty\right)$, where $\tilde{\nu}>1$ indicates reduced clearance of contrast which can be indicative of temporary retention.  

\subsection{Three Compartment Model}\label{sec:ThreeCompartmentModel}

In order to model retention of particles we consider the second compartment of the previous example as consisting of two distinct sub-compartments in addition to the first compartment, $(A,B,C)$, and allow for particle discharge from the retaining compartment $C$ into a compartment $D$. Thus, contact for particle exchange is established such that \hbox{$A\leftrightarrow B \leftrightarrow C\rightarrow D$}. Allowing for differential transport rates between all compartments (eqn.\;\ref{eqn:app_decomposerates}) and setting $k_{C\leftarrow D}=0$, or equivalently $\phi_{D}=0$, one obtains
% We can then write for both compartments:
% \begin{align*}
% 	\nu_B\,\pd{t}\phi_{B}(t) &= k_{BA}\Delta\phi_{BA}(t) + k_{BC}\Delta\phi_{BC}(t) \\
% 	\nu_C\,\pd{t}\phi_{C}(t) &= k_{CB}\Delta\phi_{CB}(t) + k_{CD}\Delta\phi_{CD}(t)
% \end{align*}
% We will allow for differential transport rates between all compartments (eqn.\;\ref{eqn:decomposerates}), and set $k_{C\leftarrow D}=0$, or equivalently $\phi_{D}=0$:
\begin{align*}
	\nu_B\,\pd{t}\phi_{B} &= k_{B\leftarrow A}\phi_A - \left(k_{B\rightarrow A}+k_{B\rightarrow C}\right)\phi_B + k_{B\leftarrow C}\phi_C \\
	\nu_C\,\pd{t}\phi_{C} &= k_{C\leftarrow B}\phi_B - \left(k_{B\leftarrow C}+k_{C\rightarrow D}\right)\phi_C
\end{align*}
Assuming that $k_{B\leftarrow C} \ll k_{B\rightarrow A} + k_{B\rightarrow C}$, i.e. the efflux from $C$ into $B$ does not substantially change the amount of \ac{CA} in $B$, the last term in the equation for $\nu_B\,\pd{t}\phi_B$ may be neglected. The response functions $R_B$ and $R_C$ can then be found to be
\begin{align*}
	R_B &= k_{B\leftarrow A}\,\e^{t/T_B}  \\
	R_C &= \dfrac{k_{B\leftarrow A}\,E_{C}}{1 - \frac{T_B}{T_C}}\left(\e^{-t/T_C} - \e^{-t/T_B} \right)
	%\nu_B\,\phi_{B} &= k_{B\leftarrow A}\,\int\limits_{0}^{t}\dd{\tau}\,\e^{\left(t-\tau\right)/T_B}\,\phi_{A}(\tau) \\
	%\nu_C\,\phi_{C} &= jj
\end{align*}
where \hbox{$T_B := \nu_B\left(k_{B\rightarrow A} + k_{B\rightarrow C}\right)^{-1}$} and \hbox{$T_C:=\nu_C\left(k_{C\rightarrow B}+k_{C\rightarrow D}\right)^{-1}$} are the \textit{average contrast transit times} in compartment $B$ and $C$, respectively, and $E_C := k_{B\rightarrow C}\left(k_{B\rightarrow A}+k_{B\rightarrow C}\right)^{-1}$ is the \textit{extraction fraction} for compartment $C$.\cite{Georgiou2017} Without additional input data one may identify $k_{C\rightarrow B}+k_{C\rightarrow D}$ as a combined efflux rate. For $k_{B\rightarrow C}=0$, $R_C = 0$ the solution becomes equivalent with the two-compartment model (sec.\;\ref{sec:TwoCompartmentModel}).

\begin{widetext}
Modeling the input function as a sum of exponentials, i.e. $\phi_A = \sum\limits_i a_i\e^{-k_i t}$, the concentration $\phi_V$ can be written as
	\begin{align}\label{eqn:3cm}
		\phi_V = \alpha\sum\limits_{i} a_i \left[ \dfrac{1-\beta}{\frac{1}{T_B} - k_i}\left( \e^{-k_i\,t} - \e^{-t/T_B} \right) + \dfrac{\beta}{\frac{1}{T_C}-k_i}\left( \e^{k_i\,t} - \e^{-t/T_C} \right) \right]
	\end{align}
where we set $\alpha=k_{B\leftarrow A}$ and $\beta = E_C / \left(1 - \frac{T_B}{T_C}\right)$ for visibility.
\end{widetext}

\subsubsection{Application to Gd-EOB-DTPA kinetics}
Applying this model to the kinetics of \ac{Eovist} in \acs{DCE}-\acs{MRI} of the liver, we can identify the compartments as follows: $\phi_A \equiv \phi_p$, $\phi_B\equiv\phi_e$, and $\phi_C\equiv\phi_h$ are the \ac{CA} concentrations in plasma, \ac{EES}, and hepatocytes, respectively. The rate constants and derived quantities follow similarly; finally $k_{C\rightarrow D}\equiv k_{b}$ is the transport rate from hepatocytes into the biliary tracts from which it is assumed the \ac{CA} will be cleared immediately.

One may correlate the rate constants $k_{h\leftarrow e}$ and $k_{h\rightarrow e}$ with expression of OATP1B1/B3 influx and efflux pumps, respectively, and the rate constant $k_{h\rightarrow b}$ with MRP2 expression.\cite{VanBeers2012} The times $T_B$ and $T_C$ then become the average extracellular and intracellular contrast transit times, respectively.\cite{Georgiou2017}

As noted previously, one may not be able differentiate between the efflux rates $k_b$ and $k_{h\rightarrow e}$, such that an increase or decrease in $k_{b}+k_{h\rightarrow e}$ may correspond to an increase in OATP1B1/B3 efflux pump, MRP2 expression, or both. Similarly, no change in that quantity compared to baseline may correspond to no change in expression, or a change in both such that change in one compensates for change in the other.

\subsubsection{Application to intracelluar distribution}

This model may similarly be applied to describe the intracellular dynamics of a contrast distribution. In this case one may identify $\phi_B\equiv\phi_{\mathrm{cytoplasm}}$ and $\phi_C\equiv\phi_{\mathrm{sub}}$ as the concentration in free cytoplasm and sub-cellular structures with prolonged retention.

% \subsection{Model verification}

% High resolution \acs{DCE}-\ac{MRI} Tofts model data were simulated with \texttt{fabber}\cite{Chappell2009} through the python application \textsl{quantiphyse}, using the population \ac{VIF} by \citeauthor{Parker2006}\cite{Parker2006} \ensuremath{K^{\mathrm{trans}}}{} and $\nu$ were varied between \qtyrange{0.1}{1.0}{\min^{-1}} and \numrange{0.1}{1.0}, respectively. Time resolution was \qty{0.1}{\min} with $300$ time points and both, clean and noisy data were generated by adding $1\sigma$ gaussian noise. Injection time was set to \qty{2}{\min} and baseline signal to \num{300}. Finally, data was simulated assuming \qty{15}{\degree} flip angle, $\ensuremath{\mathrm{TR}}=\qty{0.3}{\second}$,  $r=\qty{1.5}{(\second\,\milli M)^{-1}}$ as \ac{CA} relaxitivity, and \qty{1.3}{\second} as baseline $T_1$ value. \ac{VIF} values were calculated at the same time points.
% The \ac{3CM} was subsequently fit to both, the complete and sparsified clean and noisy data, assuming a two-exponential \ac{VIF}, and its performance compared.

\subsection{Model comparison}

The performance of the two-compartmental \ac{TM} and the proposed \ac{3CM} was compared using \acs{DCE} \acs{MRI} scans, acquired as part of routine clinical work-up, of four patients who were being treated for hepatocellular carcinoma. Scan sequences were chemical shift-based fat suppression \ac{mDIXON}.\cite{DixonThomas1984,Eggers2011} The administered \ac{CA} was \ac{Eovist}. \acs{MRI} images were renormalized using fat and muscle signals from each scan due to a lack of consistent scale parameters included in the scan's \acs{DICOM} headers.

A cylindrical ROI was placed in the abdominal aorta at the level of the liver and its intensity values averaged at each time point and fit to a sum of two exponentials.\cite{McGrath2009}
Its parameters were used to fit equation~\ref{eqn:3cm} to the whole liver averaged signal values as well as each voxel inside the liver to create 3D parameter maps.
For plotting and display, signal intensities were normalized by the sum of the maximum amplitude of the \ac{VIF}, $\Sigma_a = \sum\limits_{i}a_i$. 
%\begin{align*}
%	\phi_p(t) = a_1\,\E^{-k_1\,\Delta t}+a_2\,\E^{-k_2\,\Delta t}
%\end{align*}
%with $\Delta t = t-t_0$, where $t_0$ is the contrast onset time.

\section{Results}

\begin{figure}[tbh]
	\centering
	\includegraphics[keepaspectratio,width=0.45\textwidth]{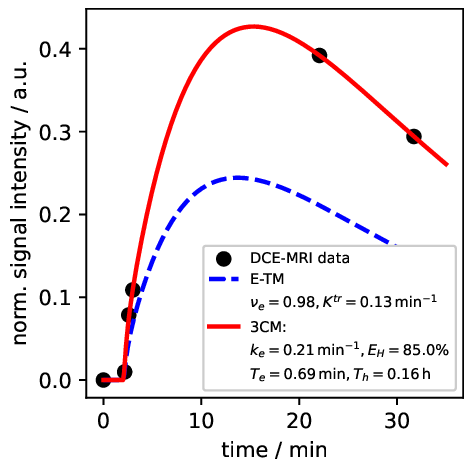}
	\caption{Model fit to DCE MRI data with very large contrast extraction fraction ($E_h>80\%$) but low mean intracellular contrast transit times ($T_h\lesssim\qty{10}{\min}$).}\label{fig:dcemri_largeEHlowTH}
\end{figure}

\begin{figure}[tbh]
	\centering
	\includegraphics[keepaspectratio,width=0.45\textwidth]{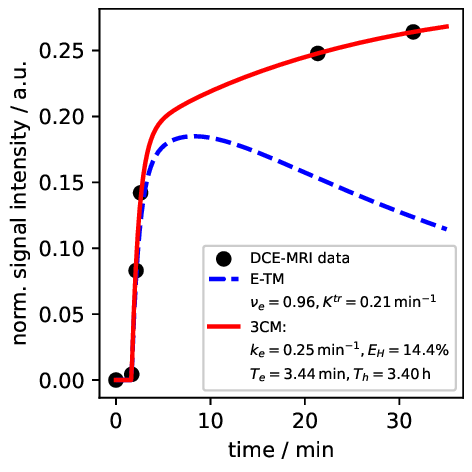}\\%
	\includegraphics[keepaspectratio,width=0.45\textwidth]{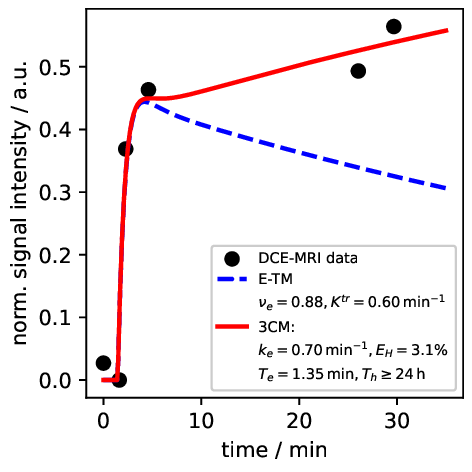}
	\caption{Model fits to DCE MRI data with long mean intracelluar contrast transit times ($T_h\gtrsim\qty{3.5}{\hour}$).}\label{fig:dcemri_longTH}
\end{figure}

Model fits to three different sets of \ac{DCE}-\ac{MRI} data are shown in figures\;\ref{fig:dcemri_largeEHlowTH} and~\ref{fig:dcemri_longTH} using both the \ac{TM} (blue-dashed) and \ac{3CM} (red-solid).
In all three cases, the \ac{TM} fits are close to the data in the early dynamic phase while showing large differences for the later delayed phase, whereas the \ac{3CM} fits are consistent with the data in all three cases.

The data shown in figure~\ref{fig:dcemri_largeEHlowTH} continually increase from \ac{CA} injection at approximately \qty{2}{\min} to $\approx 0.4\,\Sigma_a$ at approximately \qty{12}{\min} after which it decreases to $\approx 0.3\,\Sigma_a$. The corresponding \ac{3CM} model yields a high extraction fraction of $E_h=85\%$ at a $k_e\approx\qty{0.21}{\per\minute}$ and short $T_e\approx\qty{41}{\second}$ and $T_h\lesssim\qty{10}{\min}$.

In both panels of figure~\ref{fig:dcemri_longTH} the data sharply increases during the dynamic phase to approximately $0.2\,\Sigma_a$ (top) and $0.45\,\Sigma_a$ (bottom), and then continues to increase albeit at a lower rate. The corresponding \ac{3CM} models have mean intracellular contrast transit times in excess of three hours.
\ac{3CM} modeling of the data in the top panel yields a $k_e$ of \qty{0.25}{\per\minute}, a moderate extraction fraction $E_h=14.4\%$, and a mean extracellular contrast transit time $T_e\approx\qty{3.5}{\min}$ that is on the order of the duration of the dynamic \ac{DCE}-\ac{MRI} phase. Modeling of the data in the bottom panel yields a larger $k_e=\qty{0.70}{\per\minute}$ but lower $E_h=3.1\%$ and $T_e=\qty{1.35}{\min}$. The mean intracellular contrast transit times in both panels are more than five times the usual \ac{DCE}-\ac{MRI} study duration of \qtyrange{30}{45}{\min}.

\section{Discussion}

% $k_e$ being smaller and $T_e$ being larger in the right panel compared to the left panel indicate reduced transport from vasculature into \ac{EES}, which could be indicative of reduced perfusion or vessel permeability, and reduced transport from the \ac{EES} into hepatocytes, which may be due a smaller number of hepatocytes being present or under expression of the OATPB1/B3 influx pumps.

% In cases of long mean intracellular contrast transit time ($T_h\gg\qty{1}{\hour}$) one may observe the signal continuously increasing or saturating as shown in figure~\ref{fig:dcemri_modEHlongTH}, which will depend on the other three parameters and the underlying \ac{VIF}.

The \ac{TM} was found to be unable to fit the data past the dynamic phase in all cases where $E_H>0$ and $T_h>0$, i.e. when \ac{CA} is transferred into hepatocytes and subsequently retained.
In contrast, the proposed \ac{3CM} was able to model temporary retention (fig.\;\ref{fig:dcemri_largeEHlowTH}) where the mean intracelluar contrast transit time ($T_h$) is on the order or less than the duration of the \ac{DCE}-\ac{MRI} exam as well as continued uptake and retention (fig.\;\ref{fig:dcemri_longTH}) where $T_h$ is longer than the exam duration.

Due to the correlation of the parameters derived from transfer rates ($E_h$, $T_e$, $T_h$) their individual influence on the resulting model is difficult to quantify. Qualitatively however, $T_h$ determines the inflection point of the signal intensity curve past the dynamic phase and together with $T_e$ it determines the slope from \ac{CA} injection throughout the dynamic phase up to $t=T_h$.

In the setting of prolonged retention ($T_h\gg\qty{30}{\min}$), \ac{CA} transport out of hepatocytes is reduced indicating an underexpression in OATPB1/B3 efflux pumps or MRP2. For small extraction fractions, $k_h\ll k_e$, such that $T_e$ is determined primarily by $k_e$, explaining the doubling of slope in the bottom panel of figure~\ref{fig:dcemri_longTH} where $k_e=\qty{0.70}{\per\minute}$ compared to the top panel where $k_e=\qty{0.25}{\per\minute}$.
This is consistent with the \ac{TM} during the dynamic phase where the initial increase is driven by $\ensuremath{K^{\mathrm{trans}}}$.

% The influence of the individual parameters on the resulting curve can be explained using the example data shown in figures~\ref{fig:dcemri_largeEHlowTH} and~\ref{fig:dcemri_longTH}:
% \begin{itemize}
% 	\item $k_e$: 
% \end{itemize}
% Note that the \ac{TM} is unable to fit the data in all cases as it cannot account for (temporary) retention of \ac{CA}.
%; it further approaches a (transient) equilibrium with the \ac{VIF} at long times ($>\qty{10}{\minute}$) in each panel of both figures. 

It is important to note that the fit performance and validity may be limited due to the small number of data points available in the dynamic series, with \hbox{$\text{ndf}-1 = \numrange{1}{2}$}. Further investigation will prospectively acquire data with higher temporal resolution and aim to validate the fit to sparse data as presented here.
A further limitation of this study has been that no quantitative conversion between signal intensity and concentration was possible as the examinations were acquired with qualitative intent using different MRI scanners without proper $T_1$ mapping and scanner characterization. A prospective study is planned to include $T_1$ mapping to facilitate this conversion.

\section{Conclusion}

A methodology to derive multi-compartmental dynamic models has been derived and solved for two and three compartments.
The latter has been applied to sparse intensity data of \ac{DCE} \acp{MRI} employing \acs{Eovist}, highlighting the Toft's model's limitations for a hepatocyte-specific \ac{CA}.
% Beyond this, the presented method may have wide-ranging applications, including drug distribution and nanoparticle delivery and uptake.

%
% BIBLIOGRAPHY
% \section*{References}
\bibliography{library}

%aapmrev4-2.bst 2019-01-14 (MD) hand-edited version of aapmrev4-1.bst
\begin{thebibliography}{11}%
\makeatletter
\providecommand \@ifxundefined [1]{%
 \@ifx{#1\undefined}
}%
\providecommand \@ifnum [1]{%
 \ifnum #1\expandafter \@firstoftwo
 \else \expandafter \@secondoftwo
 \fi
}%
\providecommand \@ifx [1]{%
 \ifx #1\expandafter \@firstoftwo
 \else \expandafter \@secondoftwo
 \fi
}%
\providecommand \natexlab [1]{#1}%
\providecommand \enquote  [1]{``#1''}%
\providecommand \bibnamefont  [1]{#1}%
\providecommand \bibfnamefont [1]{#1}%
\providecommand \citenamefont [1]{#1}%
\providecommand \href@noop [0]{\@secondoftwo}%
\providecommand \href [0]{\begingroup \@sanitize@url \@href}%
\providecommand \@href[1]{\@@startlink{#1}\@@href}%
\providecommand \@@href[1]{\endgroup#1\@@endlink}%
\providecommand \@sanitize@url [0]{\catcode `\\12\catcode `\$12\catcode `\&12\catcode `\#12\catcode `\^12\catcode `\_12\catcode `\%12\relax}%
\providecommand \@@startlink[1]{}%
\providecommand \@@endlink[0]{}%
\providecommand \url  [0]{\begingroup\@sanitize@url \@url }%
\providecommand \@url [1]{\endgroup\@href {#1}{\urlprefix }}%
\providecommand \urlprefix  [0]{URL }%
\providecommand \Eprint [0]{\href }%
\providecommand \doibase [0]{https://doi.org/}%
\providecommand \selectlanguage [0]{\@gobble}%
\providecommand \bibinfo  [0]{\@secondoftwo}%
\providecommand \bibfield  [0]{\@secondoftwo}%
\providecommand \translation [1]{[#1]}%
\providecommand \BibitemOpen [0]{}%
\providecommand \bibitemStop [0]{}%
\providecommand \bibitemNoStop [0]{.\EOS\space}%
\providecommand \EOS [0]{\spacefactor3000\relax}%
\providecommand \BibitemShut  [1]{\csname bibitem#1\endcsname}%
\let\auto@bib@innerbib\@empty
%</preamble>
\bibitem [{\citenamefont {Tofts}\ \emph {et~al.}(1999)\citenamefont {Tofts}, \citenamefont {Brix}, \citenamefont {Buckley}, \citenamefont {Evelhoch}, \citenamefont {Henderson}, \citenamefont {Knopp}, \citenamefont {Larsson}, \citenamefont {Lee}, \citenamefont {Mayr}, \citenamefont {Parker}, \citenamefont {Port}, \citenamefont {Taylor},\ and\ \citenamefont {Weisskoff}}]{Tofts1999a}%
  \BibitemOpen
  \bibfield  {author} {\bibinfo {author} {\bibfnamefont {P.~S.}\ \bibnamefont {Tofts}}, \bibinfo {author} {\bibfnamefont {G.}~\bibnamefont {Brix}}, \bibinfo {author} {\bibfnamefont {D.~L.}\ \bibnamefont {Buckley}}, \bibinfo {author} {\bibfnamefont {J.~L.}\ \bibnamefont {Evelhoch}}, \bibinfo {author} {\bibfnamefont {E.}~\bibnamefont {Henderson}}, \bibinfo {author} {\bibfnamefont {M.~V.}\ \bibnamefont {Knopp}}, \bibinfo {author} {\bibfnamefont {H.~B.}\ \bibnamefont {Larsson}}, \bibinfo {author} {\bibfnamefont {T.-Y.}\ \bibnamefont {Lee}}, \bibinfo {author} {\bibfnamefont {N.~A.}\ \bibnamefont {Mayr}}, \bibinfo {author} {\bibfnamefont {G.~J.}\ \bibnamefont {Parker}}, \bibinfo {author} {\bibfnamefont {R.~E.}\ \bibnamefont {Port}}, \bibinfo {author} {\bibfnamefont {J.}~\bibnamefont {Taylor}},\ and\ \bibinfo {author} {\bibfnamefont {R.~M.}\ \bibnamefont {Weisskoff}},\ }\bibfield  {title} {\enquote {\bibinfo {title} {{Estimating kinetic parameters from dynamic contrast-enhanced t1-weighted MRI of a diffusable tracer: Standardized quantities and symbols}},}\ }\href {https://doi.org/10.1002/(SICI)1522-2586(199909)10:3<223::AID-JMRI2>3.0.CO;2-S} {\bibfield  {journal} {\bibinfo  {journal} {J. Magn. Reson. Imaging}\ }\textbf {\bibinfo {volume} {10}},\ \bibinfo {pages} {223--232} (\bibinfo {year} {1999})}\BibitemShut {NoStop}%
\bibitem [{\citenamefont {Sourbron}\ and\ \citenamefont {Buckley}(2011)}]{Sourbron2011}%
  \BibitemOpen
  \bibfield  {author} {\bibinfo {author} {\bibfnamefont {S.~P.}\ \bibnamefont {Sourbron}}\ and\ \bibinfo {author} {\bibfnamefont {D.~L.}\ \bibnamefont {Buckley}},\ }\bibfield  {title} {\enquote {\bibinfo {title} {{On the scope and interpretation of the Tofts models for DCE-MRI}},}\ }\href {https://doi.org/10.1002/mrm.22861} {\bibfield  {journal} {\bibinfo  {journal} {Magn. Reson. Med.}\ }\textbf {\bibinfo {volume} {66}},\ \bibinfo {pages} {735--745} (\bibinfo {year} {2011})}\BibitemShut {NoStop}%
\bibitem [{\citenamefont {Sourbron}\ and\ \citenamefont {Buckley}(2013)}]{Sourbron2013}%
  \BibitemOpen
  \bibfield  {author} {\bibinfo {author} {\bibfnamefont {S.~P.}\ \bibnamefont {Sourbron}}\ and\ \bibinfo {author} {\bibfnamefont {D.~L.}\ \bibnamefont {Buckley}},\ }\bibfield  {title} {\enquote {\bibinfo {title} {{Classic models for dynamic contrast-enhanced MRI}},}\ }\href {https://doi.org/10.1002/nbm.2940} {\bibfield  {journal} {\bibinfo  {journal} {NMR Biomed.}\ }\textbf {\bibinfo {volume} {26}},\ \bibinfo {pages} {1004--1027} (\bibinfo {year} {2013})}\BibitemShut {NoStop}%
\bibitem [{\citenamefont {Cantrell}\ \emph {et~al.}(2017)\citenamefont {Cantrell}, \citenamefont {Vakil}, \citenamefont {Jeong}, \citenamefont {Ansari},\ and\ \citenamefont {Carroll}}]{Cantrell2017}%
  \BibitemOpen
  \bibfield  {author} {\bibinfo {author} {\bibfnamefont {C.~G.}\ \bibnamefont {Cantrell}}, \bibinfo {author} {\bibfnamefont {P.}~\bibnamefont {Vakil}}, \bibinfo {author} {\bibfnamefont {Y.}~\bibnamefont {Jeong}}, \bibinfo {author} {\bibfnamefont {S.~A.}\ \bibnamefont {Ansari}},\ and\ \bibinfo {author} {\bibfnamefont {T.~J.}\ \bibnamefont {Carroll}},\ }\bibfield  {title} {\enquote {\bibinfo {title} {{Diffusion-compensated tofts model suggests contrast leakage through aneurysm wall}},}\ }\href {https://doi.org/10.1002/mrm.26607} {\bibfield  {journal} {\bibinfo  {journal} {Magn. Reson. Med.}\ }\textbf {\bibinfo {volume} {78}},\ \bibinfo {pages} {2388--2398} (\bibinfo {year} {2017})}\BibitemShut {NoStop}%
\bibitem [{\citenamefont {{Van Beers}}, \citenamefont {Pastor},\ and\ \citenamefont {Hussain}(2012)}]{VanBeers2012}%
  \BibitemOpen
  \bibfield  {author} {\bibinfo {author} {\bibfnamefont {B.~E.}\ \bibnamefont {{Van Beers}}}, \bibinfo {author} {\bibfnamefont {C.~M.}\ \bibnamefont {Pastor}},\ and\ \bibinfo {author} {\bibfnamefont {H.~K.}\ \bibnamefont {Hussain}},\ }\bibfield  {title} {\enquote {\bibinfo {title} {{Primovist, Eovist: What to expect?}}}\ }\href {https://doi.org/10.1016/j.jhep.2012.01.031} {\bibfield  {journal} {\bibinfo  {journal} {J. Hepatol.}\ }\textbf {\bibinfo {volume} {57}},\ \bibinfo {pages} {421--429} (\bibinfo {year} {2012})}\BibitemShut {NoStop}%
\bibitem [{\citenamefont {Choi}\ \emph {et~al.}(2016)\citenamefont {Choi}, \citenamefont {Huh}, \citenamefont {Woo},\ and\ \citenamefont {Kim}}]{Choi2016}%
  \BibitemOpen
  \bibfield  {author} {\bibinfo {author} {\bibfnamefont {Y.}~\bibnamefont {Choi}}, \bibinfo {author} {\bibfnamefont {J.}~\bibnamefont {Huh}}, \bibinfo {author} {\bibfnamefont {D.-C.}\ \bibnamefont {Woo}},\ and\ \bibinfo {author} {\bibfnamefont {K.~W.}\ \bibnamefont {Kim}},\ }\bibfield  {title} {\enquote {\bibinfo {title} {{Use of gadoxetate disodium for functional MRI based on its unique molecular mechanism}},}\ }\href {https://doi.org/10.1259/bjr.20150666} {\bibfield  {journal} {\bibinfo  {journal} {Br. J. Radiol.}\ }\textbf {\bibinfo {volume} {89}},\ \bibinfo {pages} {20150666} (\bibinfo {year} {2016})}\BibitemShut {NoStop}%
\bibitem [{\citenamefont {Forsgren}\ \emph {et~al.}(2014)\citenamefont {Forsgren}, \citenamefont {Leinhard}, \citenamefont {Dahlstr{\"{o}}m}, \citenamefont {Cedersund},\ and\ \citenamefont {Lundberg}}]{Forsgren2014}%
  \BibitemOpen
  \bibfield  {author} {\bibinfo {author} {\bibfnamefont {M.~F.}\ \bibnamefont {Forsgren}}, \bibinfo {author} {\bibfnamefont {O.~D.}\ \bibnamefont {Leinhard}}, \bibinfo {author} {\bibfnamefont {N.}~\bibnamefont {Dahlstr{\"{o}}m}}, \bibinfo {author} {\bibfnamefont {G.}~\bibnamefont {Cedersund}},\ and\ \bibinfo {author} {\bibfnamefont {P.}~\bibnamefont {Lundberg}},\ }\bibfield  {title} {\enquote {\bibinfo {title} {{Physiologically realistic and validated mathematical liver model revels hepatobiliary transfer rates for Gd-EOB-DTPA using human DCE-MRI data}},}\ }\href {https://doi.org/10.1371/journal.pone.0095700} {\bibfield  {journal} {\bibinfo  {journal} {PLoS One}\ }\textbf {\bibinfo {volume} {9}},\ \bibinfo {pages} {1--13} (\bibinfo {year} {2014})}\BibitemShut {NoStop}%
\bibitem [{\citenamefont {Georgiou}\ \emph {et~al.}(2017)\citenamefont {Georgiou}, \citenamefont {Penny}, \citenamefont {Nicholls}, \citenamefont {Woodhouse}, \citenamefont {Bl{\'{e}}}, \citenamefont {{Hubbard Cristinacce}},\ and\ \citenamefont {Naish}}]{Georgiou2017}%
  \BibitemOpen
  \bibfield  {author} {\bibinfo {author} {\bibfnamefont {L.}~\bibnamefont {Georgiou}}, \bibinfo {author} {\bibfnamefont {J.}~\bibnamefont {Penny}}, \bibinfo {author} {\bibfnamefont {G.}~\bibnamefont {Nicholls}}, \bibinfo {author} {\bibfnamefont {N.}~\bibnamefont {Woodhouse}}, \bibinfo {author} {\bibfnamefont {F.-X.}\ \bibnamefont {Bl{\'{e}}}}, \bibinfo {author} {\bibfnamefont {P.~L.}\ \bibnamefont {{Hubbard Cristinacce}}},\ and\ \bibinfo {author} {\bibfnamefont {J.~H.}\ \bibnamefont {Naish}},\ }\bibfield  {title} {\enquote {\bibinfo {title} {{Quantitative Assessment of Liver Function Using Gadoxetate-Enhanced Magnetic Resonance Imaging}},}\ }\href {https://doi.org/10.1097/RLI.0000000000000316} {\bibfield  {journal} {\bibinfo  {journal} {Invest. Radiol.}\ }\textbf {\bibinfo {volume} {52}},\ \bibinfo {pages} {111--119} (\bibinfo {year} {2017})}\BibitemShut {NoStop}%
\bibitem [{\citenamefont {Dixon}(1984)}]{DixonThomas1984}%
  \BibitemOpen
  \bibfield  {author} {\bibinfo {author} {\bibfnamefont {W.~T.}\ \bibnamefont {Dixon}},\ }\bibfield  {title} {\enquote {\bibinfo {title} {{Simple proton spectroscopic imaging.}}}\ }\href {https://doi.org/10.1148/radiology.153.1.6089263} {\bibfield  {journal} {\bibinfo  {journal} {Radiology}\ }\textbf {\bibinfo {volume} {153}},\ \bibinfo {pages} {189--194} (\bibinfo {year} {1984})}\BibitemShut {NoStop}%
\bibitem [{\citenamefont {Eggers}\ \emph {et~al.}(2011)\citenamefont {Eggers}, \citenamefont {Brendel}, \citenamefont {Duijndam},\ and\ \citenamefont {Herigault}}]{Eggers2011}%
  \BibitemOpen
  \bibfield  {author} {\bibinfo {author} {\bibfnamefont {H.}~\bibnamefont {Eggers}}, \bibinfo {author} {\bibfnamefont {B.}~\bibnamefont {Brendel}}, \bibinfo {author} {\bibfnamefont {A.}~\bibnamefont {Duijndam}},\ and\ \bibinfo {author} {\bibfnamefont {G.}~\bibnamefont {Herigault}},\ }\bibfield  {title} {\enquote {\bibinfo {title} {{Dual-echo Dixon imaging with flexible choice of echo times.}}}\ }\href {https://doi.org/10.1002/mrm.22578} {\bibfield  {journal} {\bibinfo  {journal} {Magn. Reson. Med.}\ }\textbf {\bibinfo {volume} {65}},\ \bibinfo {pages} {96--107} (\bibinfo {year} {2011})}\BibitemShut {NoStop}%
\bibitem [{\citenamefont {McGrath}\ \emph {et~al.}(2009)\citenamefont {McGrath}, \citenamefont {Bradley}, \citenamefont {Tessier}, \citenamefont {Lacey}, \citenamefont {Taylor},\ and\ \citenamefont {Parker}}]{McGrath2009}%
  \BibitemOpen
  \bibfield  {author} {\bibinfo {author} {\bibfnamefont {D.~M.}\ \bibnamefont {McGrath}}, \bibinfo {author} {\bibfnamefont {D.~P.}\ \bibnamefont {Bradley}}, \bibinfo {author} {\bibfnamefont {J.~L.}\ \bibnamefont {Tessier}}, \bibinfo {author} {\bibfnamefont {T.}~\bibnamefont {Lacey}}, \bibinfo {author} {\bibfnamefont {C.~J.}\ \bibnamefont {Taylor}},\ and\ \bibinfo {author} {\bibfnamefont {G.~J.}\ \bibnamefont {Parker}},\ }\bibfield  {title} {\enquote {\bibinfo {title} {{Comparison of model-based arterial input functions for dynamic contrast-enhanced MRI in tumor bearing rats}},}\ }\href {https://doi.org/10.1002/mrm.21959} {\bibfield  {journal} {\bibinfo  {journal} {Magn. Reson. Med.}\ }\textbf {\bibinfo {volume} {61}},\ \bibinfo {pages} {1173--1184} (\bibinfo {year} {2009})}\BibitemShut {NoStop}%
\end{thebibliography}%

%
% APPENDIX
\appendix
\section{Derivation}\label{sec:derivation}

The continuity equation is based on conservation of particle number or mass and relates the temporal change in particle density at a point, to the source strength of particle current density originating from that point,
\begin{align*}
\pd{t}\phi(\vec{r},t) + \div\vec{j}(\vec{r},t) &= 0
\end{align*}
Hence, \hbox{$\div\vec{j}>0$ and $\div\vec{j}<0$} denote an outflow and inflow of particles to point $\vec{r}$ while $\div\vec{j}=0$ is the static case with no net change. Closely related to the continuity equation is the diffusion equation,
\begin{align*}
\pd{t}\phi(\vec{r},t) - D\vec{\nabla}^2\phi(\vec{r},t) &= 0
\end{align*}
where the temporal change in density is related to the source strength of the density gradient field, \hbox{$\vec{\nabla}^2(\cdot)=\div\grad(\cdot)$}. Assuming diffusion processes for particle transport one can identify \hbox{$\vec{j}(\vec{r},t)=-D\,\grad\phi(\vec{r},t)$}.

To reduce the complexity of the continuity equation consider its volume integral and use Gauss's theorem:
\begin{align*}
	\int_{V}\dd{V}\pd{t}\phi(\vec{r},t) &= - \int_{V}\dd{V}\div\vec{j}(\vec{r},t) \\
	\pd{t}\Phi_{V}(t) &= -\oint_{\pd{}V}\dd{\vec{S}}\cdot\vec{j}(\vec{r},t) \\
	&= D\oint_{\pd{}V}\dd{\vec{S}}\cdot\grad\phi(\vec{r},t)
\end{align*}
Consider the integrals over distinct surfaces separately, $\oint_{\pd{}V}(\cdot)=\sum_S\int_S(\cdot)$, which may also have different diffusion coefficients \hbox{($D\equiv D(S)$)}, and identify \hbox{$\Phi_V(t)=\phi_V(t)\,V$}:
\begin{align*}
\pd{t}\phi_{V}(t)\,V &= \sum_{S}\left[D\int_{S}\dd{S}\,\hat{\vec{n}}\cdot\grad\phi(\vec{r},t)\right]
\end{align*}
If the surface normal $\hat{\vec{n}}$ and density gradient do not depend on the points on the surface the integral simplifies to
\begin{align*}
\pd{t}\phi_{V}(t)\,V &= \sum_{S,i} D\,S\,n_i\,\dfrac{\partial\phi_S(t)}{\partial x_i}
\end{align*}

Consider the one-dimensional case with $i=1$, in which $n_i=1$ and $\partial/\partial x_i\to\partial/\partial{x}$, and consider the differential in its finite form; note that $\Delta x_{ij} = -\Delta x_{ji}$ is the distance over which we consider the diffusion to take place, i.e. the membrane thickness:
\begin{align*}
\pd{t}\phi_{V}(t)\,V &= \sum_{S}\dfrac{DS}{\Delta x}\Delta\phi_S(t)
\end{align*}
The ratio of diffusion coefficient and membrane thickness can be identified as the permeability $P$ between the two compartments, while $S\equiv S_{XY}$ are the surfaces across which the diffusion takes place between compartment $Y$ and the volume of interest $X$. The product of both is the permeability surface area product,\cite{Tofts1999a} and we set \hbox{$K_{XY}\equiv K_{X\leftrightarrow Y}=\left.PS\right|_{X\leftrightarrow Y}$}. This is the volume transfer rate of particles from $X\leftrightarrow Y$:
\begin{align}\label{eqn:app_compartmentkinetics}
	\pd{t}\phi_{X}(t)\,V_{X} &= \sum_{Y}K_{XY}\Delta\phi_{XY}(t)
\end{align}
Further, define $k_{XY} := K_{XY}/V$ where $V=\sum_{i} V_{i}$ is the sum of all compartments. We shall also assume that no compartments overlap, i.e. $\forall X, Y: X\cap Y=\emptyset$.

In a departure from free diffusion, transport rates may not necessarily be symmetric between $X$ and $Y$. One may allow for this by setting
\begin{align}\label{eqn:app_decomposerates}
	K_{XY}\Delta\phi_{XY}(t) = K_{X\leftarrow Y}\,\phi_{Y} - K_{X\rightarrow Y}\,\phi_{X}
\end{align}
For example, immediate removal of particles from compartment $Y$ or a screening of those particles in $X$ will effectively set $K_{X\leftarrow Y}=0$ or $\phi_{Y}=0$ as seen from $X$.

%\subsection{General Kinetic Model}\label{sec:GeneralModel}

A \ac{VOI} may be composed of different compartments that exchange particles with one another and with an external reservoir. Measurements of signal intensity are not able to distinguish between signal emanating from one compartment over another but relate only to the overall concentration of particles in the \ac{VOI},
\begin{align}\label{eqn:app_signalpropto}
	\text{Signal} \propto \phi_{V} = \dfrac{\Phi_V}{V} = \dfrac{1}{V}\,\sum\limits_{i}\phi_i\,V_i = \sum\limits_{i}\nu_i\,\phi_i
\end{align}
Similarly, the kinetics within the \ac{VOI} are a weighted sum of the kinetics in each compartment, $\pd{t}\phi_V(t) = \sum\limits_{i}\nu_i\,\pd{t}\phi_i(t)$ where
\begin{align}\label{eqn:app_final_ode}
	\nu_i\,\pd{t}\phi_i(t) &\stackbin[(\ref{eqn:app_decomposerates})]{(\ref{eqn:app_compartmentkinetics})}{=} \sum_{j}\left[ k_{i\leftarrow j}\phi_j(t) - k_{i\rightarrow j}\phi_i(t) \right]
\end{align}

% ======================
% BODY-END
% ======================

\clearpage
\end{document}